\documentclass[showpacs,aps,prl,twocolumn]{revtex4}
\usepackage{graphicx}
\usepackage{amsmath}

\begin{document}

\title{Effects of atomic short-range order on the properties of
perovskite alloys in their morphotropic phase boundary}

\author{A.M. George$^{1}$, Jorge \'I\~niguez$^{2}$, and
L. Bellaiche$^{1}$} \address{$^1$ Physics Department, University of
Arkansas, Fayetteville, Arkansas 72701, USA\\ $^2$ Department of
Physics and Astronomy, Rutgers University, Piscataway, New Jersey
08854-8019, USA}

\begin{abstract}
The effects of atomic {\it short-range} order on the properties of
Pb(Zr$_{1-x}$Ti$_x$)O$_3$ alloy in its morphotropic phase boundary
(MPB) are predicted by combining first-principles-based methods and
annealing techniques.  Clustering is found to lead to a compositional
expansion of this boundary, while the association of unlike atoms
yields a contraction of this region.  Atomic short-range order can
thus drastically affect properties of perovskite alloys in their MPB,
by inducing phase transitions.  Microscopic mechanisms responsible for
these effects are revealed and discussed.
\end{abstract}
\date{\today}
\pacs{77.84.Dy,81.30.Bx,77.80.Bh,77.22.Ej}

\maketitle

Complex insulating perovskite alloys, with the general formula
(A$'$A$''$..)(B$'$B$''$...)O$_{3}$, are of great interest for a variety of 
device applications because of their anomalously large electromechanical responses 
\cite{Uchino,Park,Auciello}. 
Examples of such applications include piezoelectric transducers, actuators, 
as well as, dielectrics for microelectronics 
and wireless communication.

Interestingly, many perovskite solid solutions, e.g.,
Pb(Zr$_{1-x}$Ti$_{x}$)O$_{3}$ (PZT),
[Pb(Zn$_{1/3}$Nb$_{2/3}$)O$_{3}$]$_{1-x}$[PbTiO$_{3}$]$_{x}$
(PZN--PT),
[Pb(Mg$_{1/3}$Nb$_{2/3}$)O$_{3}$]$_{1-x}$[PbTiO$_{3}$]$_{x}$ (PMN--PT)
and [Pb(Sc$_{1/2}$Nb$_{1/2}$)O$_{3}$]$_{1-x}$[PbTiO$_{3}$]$_{x}$
(PSN--PT), exhibit their largest electromechanical responses for
compositions lying within the so-called morphotropic phase boundary
(MPB).  This area was thought for more than 50 years to {\it
discontinuously} separate, in the temperature-composition plane, a
rhombohedral (R) ferroelectric phase exhibiting an electric
polarization along a $\langle 111\rangle$ direction from a tetragonal
(T) ferroelectric structure having a polarization pointing along a
$\langle 001\rangle$ direction.  The recent discovery of a monoclinic
ferroelectric phase in the MPB of PZT has completely changed this
long-time accepted picture \cite{NohedaAPL}, since this new phase acts
as a structural {\it bridge} between the R and T phases
\cite{NohedaAPL,PRLPZT}.  Furthermore, the polarization in this
low-symmetry phase continuously rotates when varying the composition
\cite{PRLPZT,FE}, explaining why the MPB is the region of choice for
optimization of piezoelectric and dielectric responses
\cite{PRLPZT,FE,NohedaPRL,Fu}.  These recent findings have led to a
flurry of investigations aimed at better understanding and
characterizing the properties of the MPB in various perovskite alloys.
In particular, other low-symmetry phases have also been subsequently
discovered in PZN--PT, PMN--PT and PSN--PT (see, for instance,
Refs.~\cite{Cox1,La,Ye1,Noheda2,Kiat,Singh,submit}).

One remaining mystery of the MPB is about its 
{\it compositional width}.
More precisely, rather different widths have been reported for the same 
solid solution, depending on the growth conditions \cite{Ragini}. 
Chemical {\it short-range} ordering (SRO) is often invoked 
for this variance between different measurements \cite{Ragini}, since no 
{\it long-range} ordering occurs (to our knowledge)
in the mixed sublattice of PZT and PMN--PT near their MPB.
However, why and how SRO would affect the morphotropic phase boundary are 
two unresolved questions.
One possible reason for this lack of knowledge is that the
characterization of SRO is challenging and can only be accomplished via 
non-conventional experiment \cite{Moss}. Another plausible reason is that 
mimicking these effects --- via the use of computational schemes --- requires 
high accuracy and handling of large supercells, which are two conditions 
that are not simultaneously met by either usual first-principles techniques or 
semiempirical approaches.

In this Letter, we take advantage of the accuracy, the possibility of using large supercells, 
and the microscopic insight provided by the 
first-principles-based approach of Ref.~\cite{PRLPZT} to study the effect 
of SRO on the physical properties
of PZT in its MPB. The use of this technique (1)  proves that
SRO does have a drastic effect on these properties, especially in the Ti-poor region
of the MPB; (2) further indicates that 
short-range association of like-atoms has an opposite
effect than short-range association of unlike-atoms on the compositions delimiting the MPB;
(3) reveals the microscopic mechanisms responsible for these features.

Short-range ordering in any A(B$'_{1-x}$B$''_x$)O$_3$ 
perovskite alloy can be 
characterized by the so-called Cowley parameters, defined 
as \cite{PRB-GaAsN98}:
\begin{equation}
\alpha_{j}(x) = 1-\frac{P(j)}{x} 
\end{equation}
where ${P(j)}$ is the probability of finding a B$'$ (respectively
B$''$) atom being the $j$th nearest neighbor of a B$''$ (respectively
B$'$) atom in the mixed B-sublattice.  Then, ${\alpha_{j}(x)=0}$ for
all $j$'s represents a ${\it~ truly~ disordered~ alloy}$ while
${\alpha_{j}(x)>0}$ is associated with ${\it clustering}$, that is
$j$th nearest neighbors are preferentially of the same atomic kind.
Conversely, a configuration for which ${\alpha_{j}(x)<0}$ corresponds
to ${\it anticlustering}$, i.e. to a situation in which $j$th nearest
neighbors are preferentially of a different atomic kind.  Note that an
increase in the {\it magnitude} of ${\alpha_{j}}$ is indicative of an
increase in the {\it strength} of SRO.  We focus here on relatively
{\it small} ordering deviations occurring over the {\it first} nearest
neighbors shell from the perfectly disordered
Pb(Zr$_{1-x}$Ti$_{x}$)O$_{3}$ alloys. Practically, we limit ourselves
to the range $-0.3 <{\alpha_{1}}<0.3$, while always stipulating that
${\alpha_{2}} = {\alpha_{3}} = 0$.  We use a simulated annealing
technique \cite{PRLJorge} that allows us to determine the $12\times
12\times 12$ atomic configuration $\{{\sigma}_i\}$ --- where
$\sigma_i$=+1 or $-$1 corresponds to the presence of a Zr or Ti atom,
respectively, at the B-sublattice site $i$ of PZT --- that best
satisfies the pre-fixed Cowley parameters.  Once the desired atomic
configuration is generated, it is used for input in Monte-Carlo
simulations utilizing the first-principles-derived alloy effective
Hamiltonian proposed in Refs.~\cite{PRLPZT,FE}. All the simulations
conducted in the present study are done at 50 \,K.  The outputs of the
Monte-Carlo procedure are the set of polar local soft modes $\{ {\bf
u}_i \}$, where $i$ indexes the different 5-atom perovskite cells
\cite{footnote}, and the homogeneous strain tensor.  The $\langle{\bf
u}\rangle$ {\it supercell average} of the $\{ {\bf u}_i \}$ modes is
directly proportional to the macroscopic electrical polarization. The
{$\{ {\bf u}_i \}$ \it supercell distribution} is also used to compute
the $f_{\mu}({\bf k})$ coefficients defined as:
\begin{equation}
f_{\mu}({\bf k}) = \beta \sum_i \exp{(i{\bf k}\cdot{\bf
R}_i)}u_{i,\mu}~~ ,\label{eq:sigmak}
\end{equation}
where ${\bf k}$ is a vector in the first Brillouin zone of the simple
perovskite structure and ${\bf R}_i$ is the lattice vector associated
with cell~$i$. ${\mu}$ identifies the cartesian coordinate (the
$x$, $y$, and $z$ axes are chosen along the pseudo-cubic [100], [010],
and [001] directions, respectively) of the 
local modes ${\bf u}_i$. $\beta$ is a normalization
coefficient yielding $\sum_k |f_{\mu}({\bf k})|^{2}~=~1$.  It is important to realize
that a value close to 1 for $|f_{\mu}({\bf k})|^{2}$ at ${\bf k}={\bf 0}$ --- i.e., for the
$\Gamma$-point --- 
corresponds to the {\it homogeneous} situation where the 
$\mu$-component of ${\bf u}_i$ 
is nearly independent of the cell $i$.
On the other hand, the {\it inhomogeneous} case in which the 
local polarization fluctuates from its averaged value between
different cells yields a $|f_{\mu}({\bf k})|^{2}$ 
that is smaller than 1 for ${\bf k}={\bf 0}$. The larger this real-space 
fluctuation is, the smaller $|f_{\mu}({\bf k})|^{2}$
at $\Gamma$.

\begin{figure}[t!]
\includegraphics[width=6.8cm]{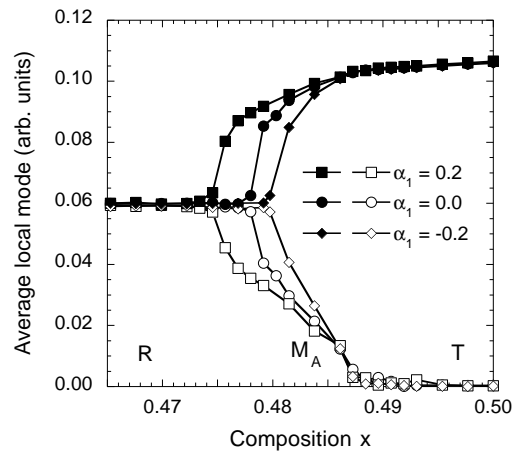}
\caption{ ($\langle u_x \rangle$, $\langle u_y \rangle$, $\langle u_z
\rangle$) Cartesian coordinates of the supercell average of the local
mode vectors, as a function of composition in
Pb(Zr$_{1-x}$Ti$_{x}$O$_{3}$) solid solutions. Circles correspond to
disordered alloys (i.e., $\alpha_1=\alpha_2=\alpha_3=0$), while
squares and losanges correspond to solid solutions exhibiting
short-range clustering ($\alpha_1=+0.2$, and $\alpha_2=\alpha_3=0$)
and anticlustering ($\alpha_1=-0.2$, and $\alpha_2=\alpha_3=0$),
respectively.  The filled symbols represent the $z$-component, while
the open symbols are associated with the $x$ and $y$ components (which
are always equal to each other).  $T$, M$_{A}$, and $R$ denote the
tetragonal, monoclinic and rhombohedral phases discussed in the text,
respectively.}
\end{figure}

Figure 1 shows the composition dependence of the $x$, $y$, and $z$
Cartesian coordinates ($\langle u_x \rangle$, $\langle u_y \rangle$,
and $\langle u_z \rangle$) of $\langle{\bf u}\rangle$ as a function of
composition in {\it disordered} materials.  As already indicated in
previous studies \cite{PRLPZT,FE}, our effective Hamiltonian approach
successfully predicts the existence of three ferroelectric phases for
random Pb(Zr$_{1-x}$Ti$_{x}$)O$_{3}$ solid solutions near its MPB: a
tetragonal (T) phase for $x>$48.7\%, a rhombohedral (R) phase for
$x<$47.7\%, and the recently discovered monoclinic \cite{NohedaAPL} --
and so-called M$_{\rm A}$ \cite{DavidMorrel} -- phase for the
composition range $\Delta x$=1.0\% in between.  The polarization is
parallel to the pseudo-cubic [001], [111], or [$vv$1] (with $0<v<1$)
direction for the T, R, and M$_{\rm A}$ phases, respectively.  The
electrical polarization can thus be viewed as rotating in a
($\bar{1}$10) plane from [001] to [111] in the M$_{\rm A}$ phase as
the Ti composition decreases~\cite{PRLPZT}.  Note that the precise
compositions at which the T-to-M$_{\rm A}$ and M$_{\rm A}$-to-R phase
transitions occur are numerically found by identifying these
compositions with those yielding a peak in the electromechanical
responses \cite{FE}.

Figure 1 also illustrates the effect of short-range clustering and
anticlustering on the compositions delimiting the MPB, by displaying
$\langle{\bf u}\rangle$ as a function of $x$ in
Pb(Zr$_{1-x}$Ti$_{x}$)O$_{3}$ alloys characterized by
$\alpha_{1}=+0.2$ and $\alpha_{1}=-0.2$, respectively.  Striking
features revealed by Fig.~1 are that SRO {\it clustering} considerably
{\it widens} this morphotropic phase boundary, while SRO {\it
anticlustering} leads to a significant compositional {\it narrowing}
of the MPB with respect to the random case.  As a matter of fact,
samples associated with $\alpha_{1}=+0.2$ and $\alpha_{1}=-0.2$
exhibit a MPB width of 1.3\% and 0.7\%, respectively. These are quite
remarkable changes with respect to the corresponding width of 1.0\%
for disordered materials, when recalling that the short-range orders
considered here only occur in the first nearest neighbor shells and
have a relatively small strength.  Interestingly, Fig.~1 shows that
the T-M$_{A}$ transition happens near the same Ti compositions of
48.7\%, independently of the value and sign of $\alpha_1$. It is thus
the M$_{A}$-R transition point that is shifting to smaller
(respectively, larger) Ti concentrations with increasing clustering
(respectively, anticlustering). In other words, SRO has nearly no
effect on the direction of the polarization in the Ti-{\it rich} area
of the MPB of PZT, while it can considerably affect the properties of
the Ti-{\it poor} region of this MPB.  This drastic difference may be
related to the fact that the {\it derivative} of $\langle{\bf
u}\rangle$ with respect to {\it composition} has a much smaller
magnitude for large Ti-compositions than for small Ti-concentrations
in the MPB of disordered PZT (see Fig.~1).  Consequently, the total
electrical polarization may be less sensitive to a slight change of
atomic configuration (i.e., a slight modification in composition or in
SRO) for larger $x$ than for smaller $x$ in the MPB of
Pb(Zr$_{1-x}$Ti$_{x}$)O$_{3}$.  The different behavior of the Ti-rich
and Ti-poor sides of the MPB may also be related to the facts that the
T-M$_{A}$ transition is second order in character while the M$_{A}$-R
transition is first order~\cite{DavidMorrel}.  Identifying the precise
microscopic mechanism responsible for the insensitivity of the Ti-{\it
rich} area of the MPB of PZT with SRO remains for future work.

\begin{figure}[ht!]
\includegraphics[width=5.9cm]{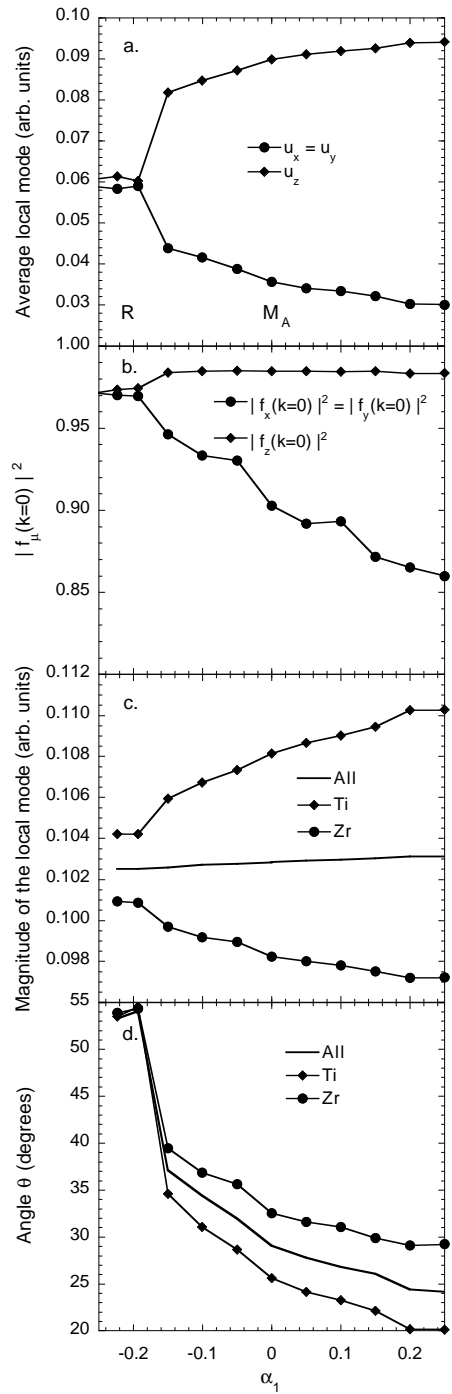}
\caption{Properties of Pb(Zr$_{0.52}$Ti$_{0.48}$)O$_{3}$ as a function
of the $\alpha_1$ SRO parameter. (a): ($\langle u_x \rangle$, $\langle
u_y \rangle$, $\langle u_z \rangle$) Cartesian coordinates of the
supercell average of the local mode vectors. (b): Fourier coefficients
defined in Eq.~(2) and computed at the $\Gamma$ point. (c): Magnitude
of the local mode vectors averaged over the 5-atom cells centered on
Zr (circles), Ti (diamonds) and on all the B-atoms (solid line).  (d):
Same as (c) but for the {\it angle} between these local-modes and the
[001] direction.}
\end{figure}

In light of the results shown in Fig.~1, 
we decided to investigate in more detail the Ti-{\it poor} area of the MPB.  
Figure~2a shows that, as ${\alpha}_1$ 
increases above
zero (clustering), the polarization of
Pb(Zr$_{0.52}$Ti$_{0.48}$)O$_3$ tends to rotate within a ($\bar{1}$10)
plane and {\it toward the [001] direction}, since $\langle u_z \rangle$ becomes larger
while $\langle u_x \rangle$ and $\langle u_y \rangle$ are reduced in magnitude. (We find that
even for very large values of $\alpha_{1}$ that are not shown here, 
the resulting phase remains monoclinic rather than becoming tetragonal.)
This is consistent
with the clustering-induced widening of the Ti-poor region of the MPB
area displayed in Fig.~1.  Figure 2a further demonstrates that, as
expected by looking at Fig.~1, decreasing ${\alpha}_1$ below
zero (anticlustering) leads to a rotation of the
polarization within a ($\bar{1}$10) plane and {\it toward the [111]
crystallographic direction}.  (In fact, the polarization actually
reaches the [111] direction characteristic of the R phase for values
of ${\alpha}_1$ below $-$0.2 in Pb(Zr$_{0.52}$Ti$_{0.48}$)O$_3$.)
Figure~2b shows the behavior of
$|f_{\mu}(\bf k=0)|^2$ (see Eq.~(2)) versus $\alpha_{1}$. 
Figure~2c shows the SRO dependence of the magnitude of 
$\langle {\bf u} \rangle$ as compared to those of
$\langle {\bf u}_{\rm Zr}\rangle$ and $\langle{\bf u}_{\rm Ti}\rangle$, the local 
mode averages restricted to cells occupied by Zr and Ti atoms, respectively.
Figure~2d displays
the variation of the {\it angle} between the pseudo-cubic [001]
direction and $\langle{\bf u}\rangle$, $\langle {\bf u}_{\rm Zr}\rangle$,
and $\langle{\bf u}_{\rm Ti}\rangle$ as a function of ${\alpha}_1$.  
Figure~2b reveals that increasing clustering leads to a larger
fluctuation of the x- and y-components of the local polarization
associated with the five-atoms-unit cells, as demonstrated by the
decrease of $|f_{x}(\bf k=0)|^2$ and $|f_{y}(\bf k=0)|^2$ as
$\alpha_{1}$ increases.  Figures~2c and~2d tell us that this large
fluctuation is correlated with the clustering-induced enhancement of
difference between $\langle {\bf u}_{\rm Zr} \rangle$ 
and $\langle {\bf u}_{\rm Ti} \rangle$: for
large values of $\alpha_{1}$, the local polarizations associated with
cells centered on Zr atoms are much closer to the [111] direction ---
and much smaller in magnitude, as consistent with 
Refs.~\cite{Egami} and~\cite{Rappe} 
--- than the local polarizations associated with
cells centered on Ti atoms.  In other words, clustering allows the
cells centered on Zr and Ti atoms to have rather different
ferroelectric properties.

Let us now discuss and understand the results depicted in Fig.~2.
Two helpful trends can be formulated by inspecting the
parameters of our alloy effective Hamiltonian for PZT \cite{PRLPZT,FE}. 
Trend~I is that alloy onsite terms  
favor {\it atomic-induced distinction} between local modes, namely 
local modes associated with cells occupied by Ti (resp. Zr) atoms 
want to align along a $\langle
100\rangle$ tetragonal (resp. $\langle 111 \rangle$ rhombohedral)
direction. Trend~II is that the intersite interactions between a local 
mode ${\bf u}_i$ centered on the $i$th cell and the local modes 
${\bf u}_j$ associated with its neighbors favor {\it similarity} between
 ${\bf u}_i$ and ${\bf u}_j$. In other words,  
an homogeneous distribution of modes is energetically favored by Trend II.
When clustering of alike atoms occurs, Ti- and Zr-rich regions tend
to develop dipole moments aligned along tetragonal and rhombohedral 
directions respectively (see Fig.~2d), in order to satisfy both
trends I and II {\it inside each of these regions}. Moreover, Trend II    
ensures that dipole moments of these different regions are correlated,
which results in a very homogeneous distribution of $u_{i,z}$ local-mode
components and a significantly disordered distribution of $u_{i,x}$ and 
$u_{i,y}$ (see Fig.~2b and note that $z$ is the preferential polarization direction of 
the Ti-rich regions).
Hence, clustering favors an overall spatially inhomogeneous 
monoclinic $M_A$ phase in the Ti-poor area of the MPB.
On the other hand, anticlustering leads to an incompatibility between 
trends~I and II even at a very short-range scale. As a matter of fact, 
local modes following their particular trend of being
either tetragonal or rhombohedral will find the opposition of their
first nearest neighbors already. This results in a more homogeneous
local-mode distribution, as shown in Fig.~2. 
Very importantly, our results prove that this homogeneity leads to 
the {\it suppression} of the monoclinic phase!  Indeed, Figs~2 show that
upon decreasing $\alpha_1$ we get a distinct transition from an inhomogeneous
to a more homogeneous distribution of local modes {\sl that is accompanied by 
a macroscopic monoclinic to rhombohedral phase transition}. It is now clear why
anticlustering results in a smaller MPB region, as shown in
Fig.~1. Our results are consistent with
Refs.~\cite{Rappe} and~\cite{OPRAPID}, in which it was pointed out that
inhomogeneity is essential to the existence of the monoclinic phase.

In summary, we have performed {\it ab-initio}-based simulations to study the effects
of SRO --- occurring 
over the first nearest neighbors' shell of the B-mixed sublattice ---
on the structural properties of PZT near its MPB.
A competition between two different energy-induced trends 
in the Ti-poor area of the MPB  
leads to a considerable widening of the MPB area when clustering occurs, while
yielding a significant narrowing of this phase boundary when anticlustering develops.
The electromechanical properties of PZT in the Ti-poor region of its MPB thus {\it drastically} depend
on the SRO, since they are extremely sensitive to the direction
of the polarization \cite{FE}.
In light of these predictions, we hope that the present Letter will encourage the attempt of experiments 
aimed at characterizing atomic short-range order in the MPB of
perovskite solid solutions. Such experiments are very frequent in metallic alloys \cite{Moss} but
(to our knowledge) have not yet been conducted in ferroelectric alloys.

This work is supported by National Science Foundation Grant
DMR-9983678 and Office of Naval Research Grants N00014-01-1-0365, 
N00014-01-1-0600 and N0014-97-1-0048.

\end{document}